\documentclass[12pt,preprint]{aastex}
\pdfoutput=1
\usepackage{amsmath}
\usepackage{graphics}      
\usepackage{graphicx}

\received{2009 Dec 07}
\slugcomment{Astrophysical Journal (Letters), Accepted: 2010 Jan 26}

\begin{document}
\newcommand{\kms}{\mbox{km~s$^{-1}$}}
\newcommand{\s}{\mbox{$''$}}
\newcommand{\mloss}{\mbox{$\dot{M}$}}
\newcommand{\my}{\mbox{$M_{\odot}$~yr$^{-1}$}}
\newcommand{\ls}{\mbox{$L_{\odot}$}}
\newcommand{\ms}{\mbox{$M_{\odot}$}}
\newcommand\mdot{$\dot{M}  $}

\title{THE ASTROSPHERE OF THE ASYMPTOTIC GIANT BRANCH STAR IRC+10216}
\author{Raghvendra Sahai}
\affil{Jet Propulsion Laboratory, MS 183-900, California Institute of Technology,
  Pasadena, CA 91109}
\email{sahai@jpl.nasa.gov}
\and 
\author{Christopher K. Chronopoulos}
\affil{Space Sciences Laboratory, University of California, Berkeley, CA 94720}

\begin{abstract}
We have discovered a very extended shock structure (i.e., with a diameter of about $24'$) 
surrounding the well-known carbon star IRC+10216 in ultraviolet images taken with the GALEX
satellite. We conclude that this structure results from the interaction of IRC+10216's molecular
wind with the interstellar medium (ISM), as it moves through the latter. All important structural
features expected from theoretical models of such interactions are identified: the
termination shock, the astrosheath, the astropause, the bowshock, and an astrotail (with vortices).
The extent of the astropause provides new lower limits to the envelope age (69,000 years) and mass
($1.4\,M_{\odot}$, for a mass-loss rate of $2\times 10^{-5}$\my). From the termination-shock
standoff
distance, we find 
that IRC+10216 is moving at a speed of about $\gtrsim$91\,\kms (1 cm$^{-3}/n_{\text{ISM}})^{1/2}$
through the local ISM.

\end{abstract}


\section{Introduction}
The carbon-rich star IRC+10216 
is the closest ($D\sim120-150$ pc)
Asymptotic Giant Branch (AGB)
star with a high mass loss rate (\mdot$\sim2\times 10^{-5}$\my) (e.g., Crosas \& Menten 1997,
Groenewegen, van der Veen, \& Matthews 1998), and has been extensively
observed from radio to optical wavelengths, and with a variety of imaging and spectroscopic
techniques. These studies have led to a fairly comprehensive model of this object, consisting of a
very cool ($\gtrsim$2000\,K)  star surrounded by a massive, extended, roughly spherical
circumstellar envelope
(CSE) expanding at 14 km~s$^{-1}$. High spatial resolution imaging observations show that
the envelope shows organized departures from spherical symmetry on the smallest scales, which are
most likely a signature that the mechanism or mechanisms which transform the round CSEs of AGB
stars into planetary nebulae (PNs) with a dazzling variety of shapes with non-spherical symmetries,
have begun their operation in this object (Sahai 2009).

Studies of IRC+10216 over the years following its discovery have led to great progress in our
understanding of the mass-loss process on the AGB, and hence the late evolution of intermediate
mass stars and their role in the enrichment of the ISM with the products of CNO and
3-$\alpha$ nucleosynthesis as well as particulate matter. But both the progenitor mass of, and the
total amount of matter ejected into the ISM by, IRC+10216, depend on the envelope's outer extent,
which remains unknown. From the deepest imaging studies undertaken so far, the
circumstellar envelope (CSE) has been traced out to a radius of $\sim200''$, via dust scattering
of Galactic starlight (Mauron \& Huggins 2000, Le{\~a}o et al.
2006). This radius corresponds to matter ejected from the star about $8000(D/120\
\text{pc})$\,yr ago.

In this Letter, we use deep GALEX 
images to trace the IRC+10216 CSE to its
outer boundary, which has become visible as a result of the CSE's interaction with the ISM,
presumably due to IRC+10216's motion through the latter. IRC+10216 is the first carbon-rich AGB
star, and the third AGB star (after R\,Hya and Mira, which are oxygen-rich: Ueta et al. 2006,
Martin et al. 2007)   
in which such an interaction has now been observed. We report on our
analysis of the shape, size and structure of the CSE-ISM interaction observed in the GALEX images,
provide new lower limits for the duration of heavy mass-loss and the total mass of ejecta in
IRC+10216, and determine its motion through the ISM.

\section{Observations \& Results}

We retrieved pipeline-calibrated FUV and NUV images of IRC+10216 from the GALEX archive; the
bandpass  (angular resolution) is 1344-1786\,\AA ($4.5{''}$) and 1771-2831\,\AA ($6.0{''}$), 
respectively, and the pixel size is $1.5{''}\times1.5{''}$ (Morrissey et al. 2005). The data
were taken on
2008 Jan 15, each with an exposure time of 8782 sec. In Fig.\,\ref{fuvnuv}a, we show a composite
FUV/NUV image, and in Fig.\,\ref{fuvnuv}b, the FUV image by itself. Field 
stars in each image have been removed using a customised IDL routine which replaces a small region
covering each star's PSF with a tile of random noise representative of the surrounding sky. The sky
noise was sampled separately at the four corners of each tile and linearly interpolated throughout,
so as to preserve gradients in the local sky background to first order.

Bright nebulosity can be seen in the center of both images, around the location of IRC+10216's
central star. In addition, the FUV image shows a bright, extended (size $\sim24'$)
ring structure which is not seen in the NUV. Very faint NUV emission is, however,
present in this region, as revealed via annular averages of the intensity (\S\,\ref{structure}).
The ring is seen mostly on the central star's
east side, and is surrounded on the outside by rather faint, diffuse NUV emission. On the west side
several additional diffuse patches of nebulosity can be seen, mostly in the FUV band, forming part
of an elongated ``tail" structure. Although the ring appears to be roughly circular around the
central star's location, closer inspection shows it be slightly flattened in the easterly
direction.

The FUV emission ring represents the interaction of the expanding CSE of IRC+10216 with
the local ISM, and the ring's East-West asymmetry implies that the star is moving roughly eastwards
through the local ISM, producing a strong shock at the eastern outer edge of
its CSE. The FUV emission is most likely not due to dust scattering of the ambient interstellar
radiation field (ISRF): the expected FUV-to-NUV brightness ratio in this case is $\sim$2.4 (since
the ISRF and dust scattering opacity at NUV wavelengths is, respectively, only about a factor 1.6
and 1.5 lower than in the
FUV (Mezger et al. 1982, Whittet 1992)) whereas the observed value is $\sim$6. Martin et al. (2007)
consider several possible mechanisms for the FUV emission in Mira,
whose wind is also known to be interacting with the ISM, and conclude that the collisional
excitation of H$_2$ by hot electrons in shocked gas is the best candidate for the FUV emission
since it produces no detectable counterpart in the NUV band. The same mechanism may be the
dominant contributor to the FUV ring emission in IRC+10216 as well, however a detailed model of
H$_2$ excitation processes is needed to properly address this issue. This exercise is outside
the scope of this Letter, and will be addressed in a follow-up paper (see \S\,\ref{structure},
Sahai \& Chronopoulos 2010).

Thus the FUV ring around IRC+10216 represents the front resulting from the shocked stellar wind
(i.e., the astrosheath); the outer edge of this ring corresponds to the astropause, and the inner
edge to the termination shock (see Fig.\,2d, Ueta 2008). The region interior to the latter
consists of the unshocked, freely-streaming stellar wind; the innermost part of this region is seen
in both the NUV and FUV images around the central star's location due to the scattering of ambient
Galactic starlight from dust in the wind. The patchy elongated astrotail shows features which
likely correspond to the vortices shed by the shock in the star's wake, seen in the numerical
simulations of a mass-losing AGB star moving through the ISM (Wareing et al. 2007). The diffuse NUV
emission around the FUV ring emission most likely represents emission from interstellar material
around the astropause. 

\section{The Astropause, Astrosheath and Bowshock}\label{structure}
\begin{figure}[htbp]
 \begin{center}
  \includegraphics[angle=90,width=12cm]{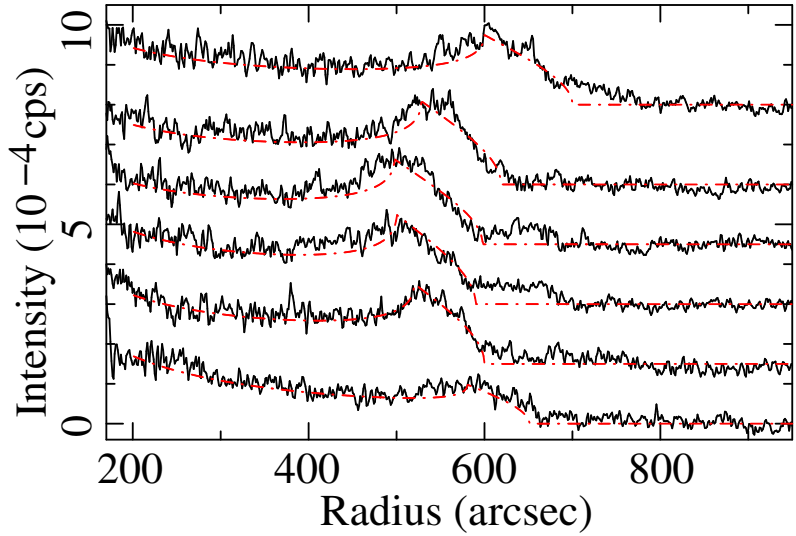}
 \end{center}
 \caption{Cuts (solid black curves) and models (dash-dot red curves) of the radial intensity 
averaged over each of six $30^{\circ}$ angular wedges on the eastern limb of the FUV emission
around IRC+10216; the intensity cuts have been shifted vertically by constant offsets for clarity.
The cut $PA$s and offsets are (from bottom to top):  $15^{\circ}(0.0), 45^{\circ}(1.5\times10^{-4}),
75^{\circ}(3.0\times10^{-4}),105^{\circ}(4.5\times10^{-4}),135^{\circ}(6.0\times10^{-4})$, and
$165^{\circ}(8.0\times10^{-4})$. Intensity units are in $10^{-4}$\,cps/pixel, implying a flux of
$1.4\times10^{-19}$ erg s$^{-1}$ cm$^{-2}$ \AA$^{-1}$ per $1.5{''}\times1.5{''}$
pixel.
}
 \label{wedgeanalysis}
\end{figure}

We have measured the astropause radius in different directions from the central star, using
radial intensity cuts from the central star location at different position angles. Since the
emission from the ring (the astrosheath) is rather faint, we have averaged the intensity over six
$30^{\circ}$ wedges spanning the eastern limb. These cuts (Fig.\,\ref{wedgeanalysis}) show that
the radius of the shock varies across the limb, reaching a minimum radius roughly in the eastward
direction, as expected from ram pressure considerations due to the (inferred) eastward motion of
IRC+10216. We have fit model radial intensities derived from a limb-brightened spherical shell to
the FUV radial brightness
profiles (assuming the surface brightness to be proportional to the column density) at each $PA$
(position angle, measured from north, towards east),
and extracted the astrosheath's inner and outer radii ($R_1$ and $R_c$, using the nomenclature in
Fig.\,1 of Weaver et al. 1977). 
A two-piece inverse-square density profile is assumed in our model, one for $160{''}\lesssim\,r<R_1$
(where the FUV and NUV intensity is well fitted by a $r^{-\alpha}$ power-law, with $\alpha$ close
to unity, implying optically-thin scattering of Galactic starlight by dust in a stellar wind 
characterized by a constant \mdot~at a constant expansion velocity), and the other for $R_c>r>R_1$,
with a jump in density at $r=R_1$. We cannot derive absolute values of the densities from our
modelling since the proportionality factor between the brightness and the column density is purely
phenomenological; furthermore, since the emission mechanisms in the two regions are different,
the value of the derived density jump is not physical. 
The values of $R_1$ and
$R_c$ which we derive are not sensitive to the assumed density profile within the astrosheath;
e.g., a model with a constant density returns very similar values, however the chi-square values of
the fits in this case are somewhat poorer. The model fits provide values of ($R_1$,\,$R_c$) in
arcseconds as follows: (580,650), (520,600), (500,590), (500,600), (530,620), (600,700) for the
cuts at $PA=15^{\circ}, 45^{\circ}, 75^{\circ}, 105^{\circ}, 135^{\circ}$, and $165^{\circ}$.

We note the presence of an FUV emission ``plateau" feature that can be seen for about 100$''$ beyond
the main peak at $r=R_c$, i.e. up to $r=R_2\sim670{''}$, in a radial cut of the FUV intensity
averaged
over a $30^{\circ}$ wedge around $PA=90^{\circ}$ (Fig.\,\ref{east-fd-nd}a). This very faint feature
can also be seen in the $30^{\circ}$-wide easterly cuts, i.e. at $PA=75^{\circ}$ and $105^{\circ}$
(Fig.\,\ref{wedgeanalysis}), and directly in the FUV image (Fig.\,\ref{east-fd-nd}b). The presence
of this sharp outer edge indicates a pile-up of gas just outside the astropause, and probably
represents the bowshock interface separating the shocked and unshocked ISM -- 
the stellar wind material is moving supersonically in the ISM. The post-shock
temperature in the bowshock region is expected to be high, about $(3/16\,k$)\
$\bar{\mu}\,V_*^2\sim10^5$K (assuming a strong shock, where the stellar velocity relative to the
ISM, $V_* = 91$\,\kms, \S\,\ref{motion}), where
$\bar{\mu}\sim10^{-24}$\,g is the mean mass per particle for fully ionized gas.
The emission in this region is most likely dominated by
emission from hot gas (i.e., a combination of continuous and line emission).

The NUV radial intensity shows a sharp rise just outside the astropause, and then a drop at the
bowshock interface, but not to zero: it extends to radii well beyond the bowshock, i.e.,
up to $\gtrsim1000{''}$, indicating that the ISM is being heated and/or excited upstream of the
bowshock. We will investigate this phenomenon and possible NUV emission mechanisms (e.g.,
continuous emission from hot gas, line emission from the excited products of ISM ions
charge-exchanging with energetic neutrals from the stellar wind) in a follow-up paper (Sahai \&
Chronopoulos 2010, in prep). Here, we merely note that an analogous situation is found for the
heliosphere, where heating and compression of the local ISM (LISM) occurs upstream of the bowshock
by charge exchange between secondary H atoms in the solar wind and protons from the LISM plasma
(Izmodenov 2004). We also note that at the termination shock, $R_1$, the NUV
intensity shows an abrupt departure (upwards) from its power-law decrease seen at smaller radii
(the latter results from dust scattering in the unshocked wind, see \S\,\ref{motion}), implying
that there are additional emission mechanisms operational in the astrosheath region than just H$_2$
line emission (which only contributes in the FUV band). 

\section{IRC+10216's Motion through the ISM, Mass-Loss Duration and Circumstellar
Mass}\label{motion}
We estimate the star's velocity $V_*$ through the surrounding ISM using the relationship between
$l_1$, the distance of the termination shock from the star along the astropause's symmetry axis 
(i.e., the termination-shock standoff distance), and $V_{*} (\kms)=10\,V_{*,6}$
(Eqn. 1 of van Buren \& McCray 1988):
\begin{equation} l_1 (cm) =1.74\times10^{19}\,(\dot{M}_{*,-6} V_{w,8}\,/\
\bar{\mu}_H\,n_{\text{ISM}})^{1/2}\,V_{*,6}^{-1}
\label{vanburen} 
\end{equation}

\noindent
where $\dot{M}_{*,-6}$ is the stellar mass-loss rate in units of $10^{-6}$\,\my, $V_{w,8}$ is
the wind velocity in units of $10^3$\,\kms, $\bar{\mu}_H$ is the dimensionless mean molecular mass
per H atom, and $n_{\text{ISM}}$ is the ISM number density in cm$^{-3}$.

Given the strong asymmetry between the eastern and western hemispheres, we first make the
simplifying  
assumption that the astropause's symmetry axis lies in the sky-plane, i.e., the inclination
angle, $\phi=90^{\circ}$. 
We find $l_1=R_1=8.6\times10^{17}$\,cm, using the value of
$R_1=478''$ as derived from the easterly cut (shown in Fig.\,\ref{east-fd-nd}). Substituting this
value of $l_1$ in  
Eqn.~\ref{vanburen}, with $\dot{M}_{*,-6}=20$, $V_{w,8} =
0.014$, $\bar{\mu}_H = 1.33$ (for an 89/11 mixture of hydrogen/helium), and
$n_{\text{ISM}}=1$, we get $V_* = 91$\,\kms. Note that the value of $V_*$ (i) does not
depend on the poorly known distance, $D$, to IRC+10216, since both $l_1$ and
$\dot{M}_{*,-6}^{1/2}$ scale
linearly with $D$, and (ii) depends only weakly on the uncertain value of the ISM density at
IRC+10216's location. 

\begin{figure}[htbp]
  \vskip -5in
\begin{center}
\includegraphics[width=16cm]{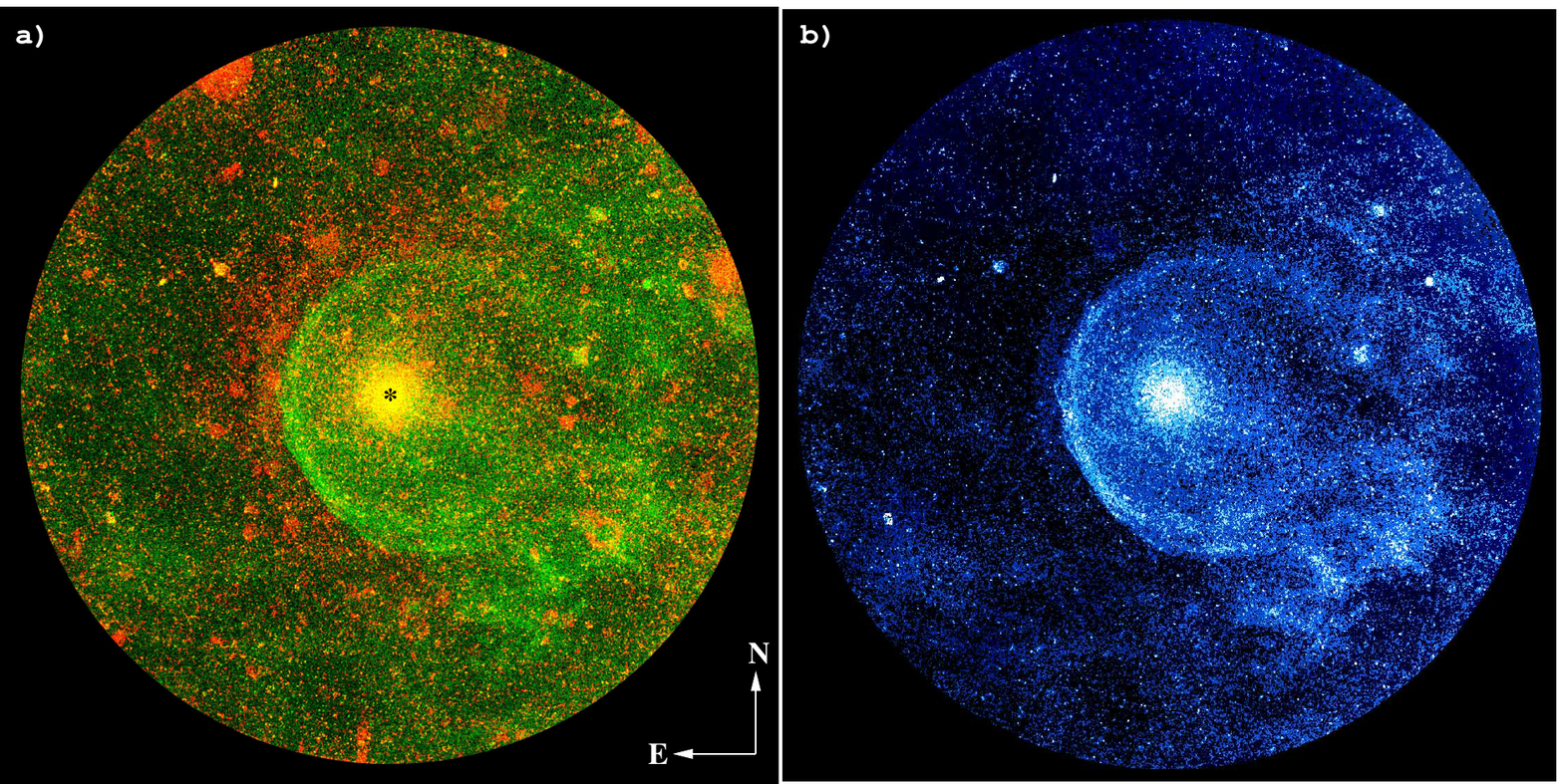}
\end{center}
\caption{(a) Composite [NUV ({\it red}) \& FUV ({\it green})] GALEX image of IRC+10216 (the circular
field-of-view (FOV) has a diameter of 61.6'$\times$61.6'); the NUV\,(FUV) image was boxcar-smoothed
using a $3\times3$ ($2\times2$) pixel box, and displayed using a linear (square-root) stretch.  The
location of the central star is indicated by a $\star$; the bright round red patches and streaks at
the edges of the NUV image are due to bright stars which could not be removed, and detector edge
artifacts. (b) The FUV image (same FOV as in a), which is less affected by bright star residuals
and artifacts, boxcar-smoothed using a $3\times3$ pixel box, and displayed using a linear stretch
(in false color), to clearly show the detailed structure of the astropause and its tail.
}
\label{fuvnuv}
\end{figure}

\begin{figure}[htbp]
  \vskip -5in
 \begin{center}
  \includegraphics[width=16cm]{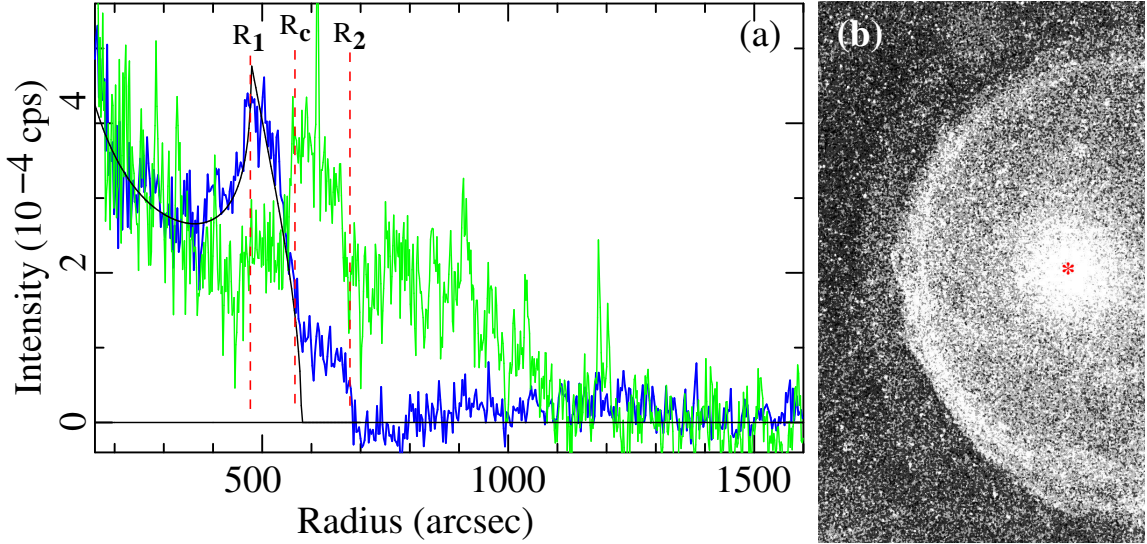}
 \end{center}
 \caption{({\it a}) Cuts of the the NUV ({\it green}) and FUV ({\it blue}) radial intensities
obtained by averaging over a $30^{\circ}$ wedge around
$PA=90^{\circ}$. A model fit to the FUV intensity, of a limb-brightened shell ({\it black}), as well
as approximate locations of the termination shock ($R_1$), the astropause ($R_c$) and the bowshock
($R_2$), are also shown. The FUV intensity (units as in Fig.\,\ref{wedgeanalysis}) has been
scaled up by a factor 2. Intensity units for the NUV are in $10^{-4}$\,cps/pixel, implying a 
flux of $2.06\times10^{-20}$ erg s$^{-1}$ cm$^{-2}$ \AA$^{-1}$ per $1.5{''}\times1.5{''}$ pixel.  
({\it b}) Part of the FUV emission around IRC+10216 (box size is 17.5'$\times$29'; orientation,
smoothing and stretch as in Fig.\,\ref{fuvnuv}). The location of the central star is indicated by a
$\star$.
}
 \label{east-fd-nd}
\end{figure}

In order to estimate the inclination angle accurately, we will need to fit a 3-D model of a
paraboloidal-shaped emitting astrosheath to the observed emission, which is outside the scope of
this Letter.  Mac Low et al.'s (1991: MLetal91) 
paraboloidal bowshock models for various inclination angles (their Fig.\,5)
show that as $\phi$ becomes {\it smaller}, the ratio of the radial distance between the
star and the
apex of the projected emission paraboloid, to the (unprojected) standoff distance, becomes
{\it larger}. A visual comparison of IRC+10216's FUV emission morphology (Fig.\,\ref{fuvnuv})
with the surface brightness countours in Fig.\,5 of MLetal91, indicate that $\phi$ could be small,
$\sim30^{\circ}$, in which case our measured value of $R_1$ is larger than $l_1$ by a significant
factor. We speculate that this factor may be as large as $\sim1.5-2$, by comparing the length of
the (unprojected) standoff distance vector, to the distance between the star and the apex as
defined by the emission contours for $\phi=30^{\circ}$ and $90^{\circ}$ in Fig.\,5 of MLetal91.
Thus the value of $V_*$ may be as high as $\sim160$\,\kms.

The FUV emission traces the spherical AGB stellar wind to a much larger distance from the star than
previous measurements (200$''$, Le{\~a}o et al. 2006). For example, the radial extent of the
astropause along directions orthogonal to the direction of motion of the central star, i.e., 
at positions angles, $PA=0^{\circ}$ and $180^{\circ}$, is $700''$ (84,000 AU at $D=120$\,pc).
Hence, we can use the astropause size to substantially revise (upwards) previous estimates of the
duration, $P$, of heavy mass-loss in IRC+10216 (8000 yr, Le{\~a}o et al. 2006). We estimate $P$  
by deriving expansion time-scales ($P_u, P_s$) for the unshocked and shocked wind regions  
separately; $P_u=19,480$\,yr from the ratio of the termination shock radius ($478{''}$) to
$V_w$, and $P_s=49,740$\,yr from the ratio of the astrosheath width ($102{''}$) along the symmetry
axis (since the velocity vectors are radial along this axis) to an average velocity for this
region, $\bar{V_s}=1.17\,\kms$. We take $\bar{V_s}=V_s/2$, where $V_s=V_w\
(\gamma-1)/(\gamma+1)=V_w/6=2.33\,\kms$, is the velocity in the astrosheath just beyond the
termination shock, with $\gamma=7/5$ for diatomic gas, and assuming the latter to be adiabatic. The
actual value of $V_s$ should be less
than the adiabatic value, since the astrosheath appears to have cooled to some degree (the
astrosheath's average width of 100$''$, or $1.8\times10^{17}$ cm, is smaller than the adiabatic
value,  $\approx0.47\,l_1=4\times10^{17}$ (Eqn. 2, Van Buren \& McCray 1988)). 
Furthermore, once a complete balance has been established
between the ram pressures of the stellar wind and the ISM, the leading edge of the astropause
remains a fixed distance ahead of the moving star (Weaver et al. 1977).  Hence,  
$P=P_u+P_s=69,220$\,yr is a lower limit, and we conclude that IRC+10216
has been undergoing mass-loss for at least 69,000\,years, and the total CSE mass is
$>1.4\,M_{\odot}$.  

In summary, the GALEX
images of IRC+10216 show an unprecedented detailed picture of the interaction of the wind from an
AGB star with the ISM, due to its motion in the latter, allowing us to 
identify all the important structural features expected from theoretical models of such
interactions. This interaction process has been
observed in the past for other AGB stars (e.g., R Hya: Ueta et al. 2006, Mira: Martin et al. 2007). 
It is noteworthy that the shock structure on the east side of IRC+10216 is relatively smooth and
well-defined, and does not show the large-scale instabilities (of the order of the standoff
distance) 
which Blondin \& Koerwer (1998) predict for stars with slow, dense winds moving relative to the ISM
at high speeds (of order 60\,\kms). Our study provides valuable new data for understanding the
CSE-ISM interaction, especially since the mass-loss rate in IRC+10216 is about 2 orders of
magnitude greater than in R Hya and Mira. New hydrodynamic simulations of stellar wind-ISM
interactions (such as, e.g., by Wareing, Zijlstra \& O'Brien 2007) to explore the high mass-loss
regime represented by IRC+10216's stellar wind should be carried out.

\acknowledgments
We thank Drs. D. van Buren, M. Mac Low, T. Ueta, C. Wareing and H.-R. M{\"u}ller for helpful
discussions. 

\clearpage

\end{document}